\documentclass[sn-mathphys]{sn-jnl}

\jyear{2023}%

\theoremstyle{thmstyleone}%
\theoremstyle{thmstyletwo}%

\theoremstyle{thmstylethree}%

\usepackage{comment}
\usepackage{amsmath}
\usepackage{booktabs}
\usepackage{siunitx}
\usepackage{lineno}
\usepackage{hyperref}
\usepackage{pdfpages}
\DeclareMathOperator{\sgn}{sgn}
\DeclareMathOperator{\Arg}{Arg}

\begin{document}

\title[ ]{High-Fidelity and High-Speed Wavefront Shaping by Leveraging Complex Media}


\author[1]{\fnm{Li-Yu} \sur{Yu}}\email{lyyu@mit.edu}

\author*[1]{\fnm{Sixian} \sur{You}}\email{sixian@mit.edu}

\affil*[1]{\orgdiv{Department of Electrical Engineering and Computer Science}, \orgname{Massachusetts Institute of Technology}, \orgaddress{\street{50 Vassar St.}, \city{Cambridge}, \postcode{02142}, \state{Massachusetts}, \country{USA}}}


\abstract{
Achieving high-precision light manipulation is crucial for delivering information through complex media with high fidelity. However, existing spatial light modulation devices face a fundamental tradeoff between speed and accuracy. Digital micromirror devices (DMDs) have emerged as a promising candidate as accessible high-speed wavefront shaping devices but at the cost of compromised fidelity, largely due to the limited control degrees of freedom and the challenge of numerically optimizing a binary amplitude mask.
Here we leverage the sparse-to-random transformation through complex media to overcome the dimensionality limitation of spatial light modulation devices. We demonstrate that pattern compression in the form of sparsity-constrained wavefront optimization allows sparse and robust wavefront representations of generic patterns in the random basis provided by the complex media, and thus effectively addresses the dimensionality limitation of DMDs, which significantly improves the projection fidelity without sacrificing the full frame rate (22\,kHz), hardware complexity, or optimization time (0.5\,s for 1000 frames). Since the dimensionality limitation is intrinsic to spatial light modulation devices and sparse-to-random transformation to complex media, our methods can be generalized to different pattern types, complex media, and device settings, supporting consistent superior performance across different types of complex media with up to an 89\% increase in projection accuracy and a 126\% improvement in speckle suppression. The proposed optimization framework has the potential to enhance existing holographic setups without any change to the hardware, enable high-fidelity and high-speed wavefront shaping through different scattering media and platforms, and directly facilitate a wide range of physics and real-world applications.}



\maketitle
\section{Introduction}\label{sec1}
Light scattering is ubiquitous in fog, biological tissues, and other complex media with inhomogeneous and disordered structures, which prohibits direct access to the scene beyond a short transport mean free path, e.g., \SI{100}{\micro\meter} in biological tissues \cite{Ntziachristos2010, Mosk2012, Cao2022, Gigan_2022}. Over the past two decades, precise manipulation of light has been demonstrated in and through various complex media, promising a wide range of applications in microendoscopy \cite{Choi2012, Loterie:15, Turtaev2018, Ohayon:18, Amitonova2020, Lee2020}, non-invasive deep-tissue imaging \cite{Vellekoop:07, Popoff2010-2, Hsieh:10-2, Katz2011, Si2012, Horstmeyer2015, Boniface2020, Yoon2020, Aizik:22}, holographic optical tweezers \cite{Horodynski2020, Leite2018}, microfabrication \cite{Morales-Delgado:17, Konstantinou2023}, and optical telecommunications \cite{Richardson2013, Ruan2021}. The rapid progress in wavefront shaping in complex media can be partly attributed to the increasing availability and performance of spatial light modulation devices such as liquid-crystal-based spatial light modulators (LC-SLMs) and digital micromirror devices (DMDs). These devices compensate for the scattering process by generating conjugated light fields through transmission matrix (TM) inversion \cite{Vellekoop:07, Popoff2010, Choi2012, Loterie:15, Boniface2020, Pai2021, Lee2022_trmatrix, Bender2022_2}, optical phase conjugation (time reversal) \cite{Yaqoob2008, Hsieh:10-2, Ma2014, Feldkhun:19, Cheng2023}, or iterative wavefront optimization \cite{Nixon2013, Lai2015, Horstmeyer2015, Yeminy2021, Cheng:22, Aizik:22}. For applications that require real-time speed and high-precision light manipulation, such as holographic optogenetics \cite{Pegard2017, Ruan2017}, multimode-fiber-based endoscopy \cite{Choi2012, Loterie:15, Turtaev2018, Ohayon:18, Amitonova2020, Lee2020}, and holographic 3D printing \cite{Morales-Delgado:17, Konstantinou2023}, high-speed and high-fidelity wavefront shaping through complex media is in high demand for fast and precise projection of optimized light fields. 

However, almost any existing spatial light modulation devices have a fundamental tradeoff between speed and accuracy due to hardware limitations, including data transfer rates, driving voltages, and heat dissipation. This tradeoff between speed and accuracy is evinced by the competition between the frame rate and the modulation depth in the spatial light modulation devices, resulting in dimension limitation in wavefront shaping problems. For example, LC-SLMs feature high-precision (8-12 bits) phase modulation and have been demonstrated for high-fidelity wavefront shaping in various complex media \cite{Rahmani2020, Ploschner:15, Loterie:15, Boniface2020, Pai2021}, yet the frame rates are limited to 50-600 Hz. While a 350 kHz 1D SLM has been demonstrated for wavefront shaping in complex media \cite{Tzang2019}, its high frame rate is at the cost of a total 1088 degrees of freedom, which limits the enhancement ratio and, consequently, the focusing quality. On the other hand, DMDs can achieve a frame rate of up to 22 kHz enabled by a high-speed micro-electro-mechanical system, while the precision can be unsatisfactory for high-fidelity wavefront shaping due to the limited modulation depth (1-2 bits) in amplitude. To enable high-precision wavefront shaping in high-speed applications, a wide variety of approaches, including the Lee hologram method \cite{Lee:74, Conkey:12}, the superpixel method \cite{Goorden:14}, and the island algorithm \cite{Zamkotsian:19}, have been proposed to convert a binary DMD pattern into a complex wavefront, which is commonly used in TM-based approaches.

Despite the development of holographic coding schemes for DMDs, high-speed, high-fidelity wavefront shaping in complex media remains challenging using DMDs due to the limited degrees of freedom and the difficulty of optimizing a binary amplitude mask. Considering the large number of pixels involved, typically ranging from hundreds of thousands to four million, directly performing binary optimization \cite{Akbulut:11, Zhang_2014, Woo:21} of the binary mask (which is non-differentiable) for the entire DMD frame can lead to significant computational complexity and convergence to suboptimal solutions. To address the artifacts arising from binary modulation, end-to-end \cite{Lee2022_dmd} and deep-learning-based methods \cite{HosseinEybposh:20} were proposed to take advantage of scientific understanding (mathematical models) or observations (training datasets) of a specific system to accommodate the underlying artifacts. However, these methods either require precise calibration of a specific forward model corresponding to a predefined system configuration or substantial training datasets from specific types of complex media, which makes these methods system- and data-dependent, limiting their generalizability to new systems and complex media platforms. Towards a more system- and data-agnostic approach, temporal multiplexing methods \cite{BoonzajerFlaes:21, Ayoub2021} exploit the statistical properties of speckle patterns to improve the projection quality via temporal averaging, but at the cost of a roughly ten-fold reduction in frame rate. An alternative approach, that is also system- and data-agnostic yet at the full frame rate, is the utilization of the phase-only constraint in DMD-based wavefront shaping techniques. This method involves optimizing the wavefront using only the phase information, while the amplitude is set to a constant value. Optimizing the wavefront with a phase-only constraint is one of the most popular and successful methods in computer-generated holography, microscopy, and wavefront shaping in complex media using phase-only SLMs. While this approach has been successfully demonstrated on DMDs \cite{Conkey:12, Turtaev2018, Ohayon:18}, it omits the possibility of simultaneous amplitude and phase modulation \cite{Mirhosseini:13, Georgieva2022, Gomes:22} and could be susceptible to ill-posedness and ill-conditionedness of the inverse problem without proper regularization \cite{Lee2020, Tuckova:21}, leading to suboptimal inverse solutions with limited projection fidelity.

\begin{figure*}[h]
\centering
\includegraphics[width=0.95\textwidth]{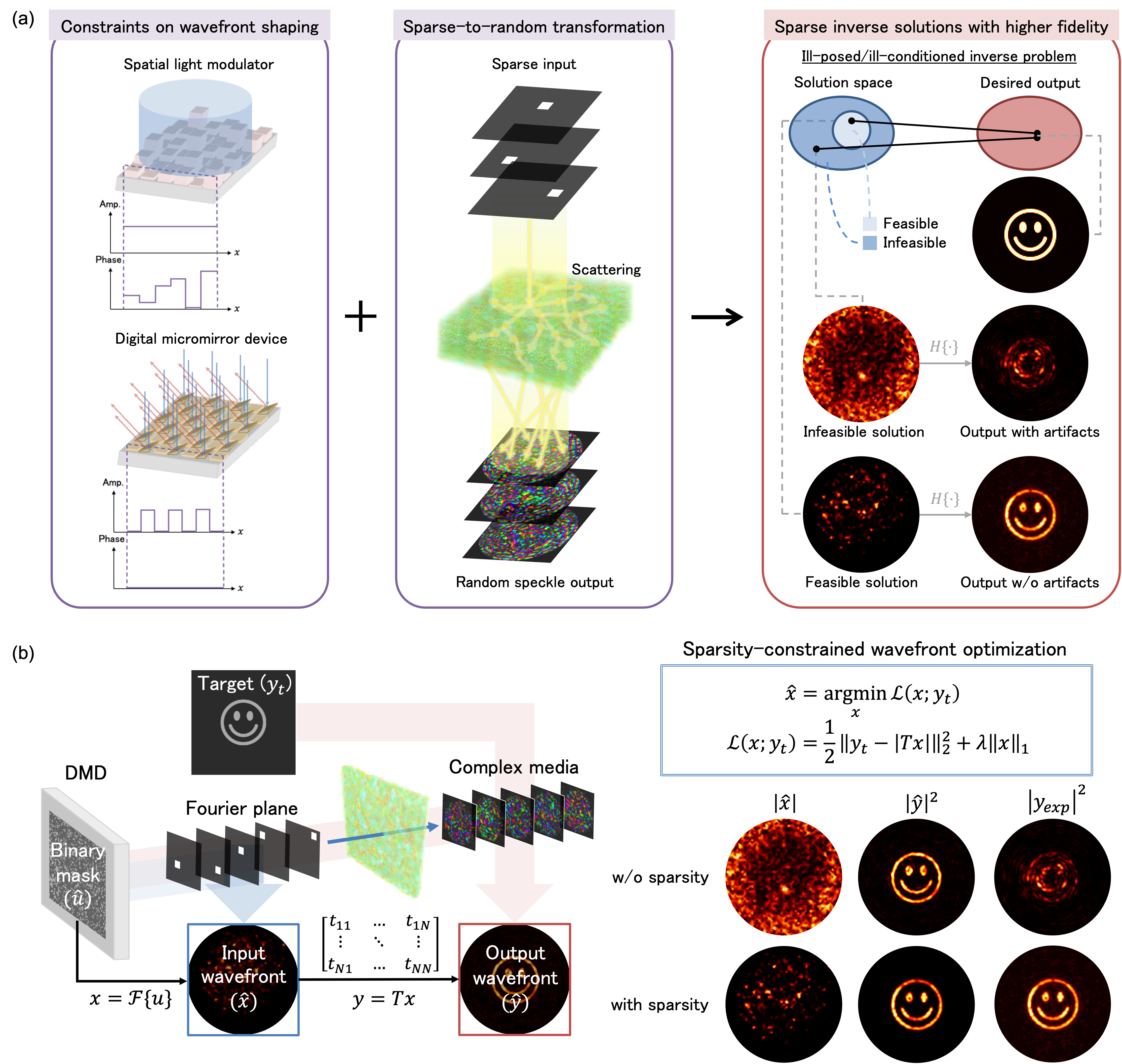}
\caption{\textbf{Design principle of sparsity-constrained light shaping through complex media.} (a) The challenge of achieving arbitrary light manipulation in complex media can be addressed by recognizing and leveraging two physical properties in the optimization framework: the dimensionality limitation of wavefront shaping and the sparse-to-random transformation of complex media. (b) The wavefront optimization problem in complex media involves optimizing the pattern displayed on a spatial light modulation device to generate a given target pattern in or through a scattering medium. To achieve high-fidelity pattern projection at a full frame rate, a sparsity constraint in the Fourier plane of the DMD is introduced through $l_1$ regularization as a physics prior, which leverages the fact that patterns are compressible in a random basis provided by complex media to overcome the limited control degrees of freedom.}\label{figure1}
\end{figure*}

To fill this gap, we leverage the intrinsic random multiplexing in complex media to remedy the dimensionality limitation problem of spatial light modulation devices, allowing for a sparse and robust wavefront representation to achieve high-fidelity projection through complex media at a full DMD frame rate. We propose a sparsity-constrained optimization framework that accounts for two physical properties: 1) the limited degrees of freedom of spatial light modulation devices, and 2) the sparse-to-random transformation caused by the light scattering in complex media (Fig. \ref{figure1}(a)). Towards the goal of an accessible, generalizable, high-speed, high-fidelity projection method, we investigate the underlying limitations of DMD-based wavefront shaping and the sparse representations of wavefronts in a random basis provided by complex media, yielding a new optimization framework that simultaneously targets high physical feasibility and numerical optimality (Fig. \ref{figure1}(b) and Methods \ref{subsec-optimization}). By incorporating the dimensionality limitation through $l_1$ regularization, our approach demonstrates consistently higher-fidelity projections across different types of complex media, showing an up to an 89\% increase in projection accuracy and a 126\% improvement in speckle suppression through graded-index multimode fibers, step-index multimode fibers, and diffusers at the speed of 22 kHz. 

\section{Results}\label{sec2}
\subsection{Sparsity constraint for DMD-based wavefront shaping}\label{subsec1}

\begin{figure*}[]
\centering
\includegraphics[width=0.95\textwidth]{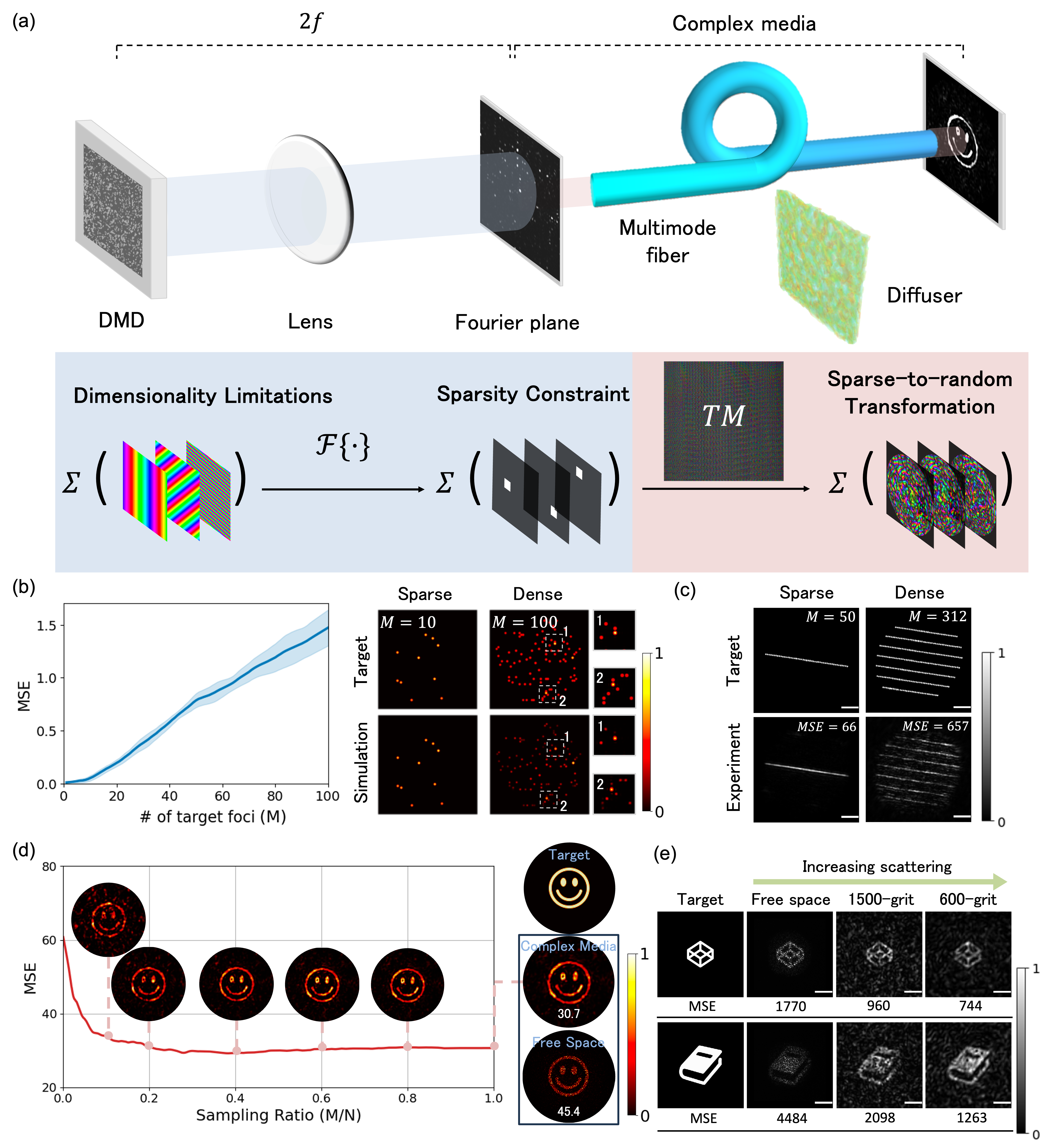}
\caption{\textbf{Numerical and experimental illustration of the sparsity constraint of wavefront shaping and the sparse-to-random transformation of complex media.} (a) Schematics illustrating sparsity-constrained DMD-based wavefront shaping. Without scattering, the set of attainable wavefronts is limited by the degrees of freedom of the DMD, which can be expressed as a sparsity constraint in the Fourier domain. However, this dimensionality limitation can be overcome by the sparse-to-random transformation through a complex medium, which supports a sparse representation of a generic pattern in a random basis. (b) Simulation of wavefront shaping in the Fourier plane without scattering. The wavefront mean squared error (MSE) increases with a larger number of foci. Ten different random distributions of foci are simulated, and the resulting MSE is averaged. (c) Experimental results of different pattern sparsity in the Fourier plane using the Lee hologram method. (d) Simulation of pattern reconstruction with different sampling ratios (the fraction of the number of input modes) through a diffuser. (e) Experimental results of wavefront shaping in media with different levels of scattering strength, including free space and a diffuser with 1500 and 600 grit polishes (DG10-1500 and DG10-600, Thorlabs, respectively). The scale bars are \SI{10}{\micro\meter}}\label{figure2}
\end{figure*}

In experiments, we observe that DMDs show higher fidelity when the targets are sparser in the Fourier plane due to the limited degrees of freedom (Fig. \ref{figure2}(c) and Supplementary Figure S1). To gain quantitative insight into the relationship between the wavefront fidelity and the sparsity in the Fourier plane, we conduct a simulation and an experiment (see Methods \ref{subsec-simulation} for implementation details) of DMD-based wavefront shaping using the Lee hologram method in a Fourier domain setup without complex media similar to Fourier transform holography \cite{Stroke1965} (Fig. \ref{figure2}(a)). The simulation result in Fig. \ref{figure2}(b) illustrates that projecting more foci in the Fourier plane simultaneously leads to a decreased projection quality, which is in agreement with the experimental result shown in Fig. \ref{figure2}(c) and Supplementary Figure S1. Supplementary Note 1 details a simplified theoretical explanation for our experimental observations by deriving the wavefront error as a result of the limited control degrees of freedom of DMDs. These observations and analyses demonstrate the intrinsic tradeoff between fidelity and the complexity (bandwidth, sparsity) of the pattern projected by DMDs. Due to the preference of sparse patterns, such dimensionality limitation of DMD-based wavefront shaping can be potentially described by the sparsity of the wavefront in the Fourier plane. Compared to phase-only and binary constraints, incorporating the intrinsic sparsity constraint of the hardware as a $l_1$ minimization  offers a balance between experimental projection fidelity and numerical optimality. It allows wavefronts with non-uniform amplitude distributions, which could be a more practical solution for DMDs. In addition, the sparsity constraint in the form of $l_1$ minimization can potentially converge better and faster, which will be important for noisy measurements and real-time applications as we will discuss in Section \ref{subsec3}. Now that we gain quantitative and physical insights into the sparsity constraints for DMD-based wavefront shaping, next, we investigate how to build on this insight in the optimization framework to enable high-fidelity high-speed light manipulation through complex media in the following sections. 

\subsection{Sparse-to-random transformation via complex media}\label{subsec1-2}
Given the observation that higher fidelity is associated with lower bandwidth (sparse) targets (Fig. \ref{figure2}(b-c) and Supplementary Figure S1), we seek a way to convert the high-bandwidth signals (arbitrary, generic patterns) to low-bandwidth ones to accommodate the sparsity constraint of DMDs. We find that, rather than being detrimental to precise wavefront shaping, complex media can support sparse representations of generic patterns in a random basis defined by a set of speckle patterns, which we refer to as sparse-to-random transformation in this article. This property, together with $l_1$ minimization, allows the recovery of a generic pattern through scattering by a relatively sparse and robust wavefront at the input end, which overcomes the hardware sparsity constraint of DMDs (Fig. \ref{figure2}(d-e)). These key insights are consistent with the theory of compressive sensing (incoherence and random sensing \cite{Candes2006, Candes2008}) and have been supported by the numerical and experimental observations (Fig. \ref{figure2}(d-e) and Supplementary Figure S2) that patterns are compressible in random basis provided by scattering media (more examples shown in Supplementary Figure S2). In light of the synergy between the two key insights, we propose a new solution – pattern compression in the form of sparsity-constrained wavefront optimization (Fig. \ref{figure1}(b) and Methods \ref{subsec-optimization}).

Consistent with the results in Fig. \ref{figure2}(b-c), without the complex media, the simulated projection quality is poor due to the dimensionality limitation of DMDs shown in the simulation in Fig. \ref{figure2}(d). However, using the same number of input modes in the Fourier domain, the light shaping fidelity is significantly improved with the addition of complex media (sampling ratio $M/N = 1.0$), which is experimentally validated in Fig. \ref{figure2}(e). Moreover, this sparse-to-random transformation is more pronounced in complex media with stronger scattering, leading to considerable enhancements in the quality of the dense pattern projection (Fig. \ref{figure2}(e) and Supplementary Figure S4). To investigate the effectiveness of a sparse input in generating generic images, we progressively decrease the sampling ratio, which is defined as the fraction of input modes utilized in the image reconstruction, and compared the performance with the reference image directly generated from the inverse solution. Fig. \ref{figure2}(d) shows that reduction of input modes does not necessarily lead to reduction of projection accuracy. Interestingly, a relatively low sampling ratio is sufficient to yield an image of comparable quality to the reference image (around 0.25 in this example). A few more numerical examples of pattern compression in the random basis provided by complex media are shown in Supplementary Figure S2. These numerical and experimental results show that 1) most patterns are compressible using a random basis provided by the complex media, and 2) the sparse-to-random transformation of the complex media, together with  $l_1$ minimization, allows sparse reconstruction of generic patterns, which conveniently compensates for the dimensionality limitation of DMDs.

\subsection{High-fidelity light shaping via sparsity-constrained optimization}\label{subsec3}
\begin{figure*}[h]
\centering
\includegraphics[width=1.0\textwidth]{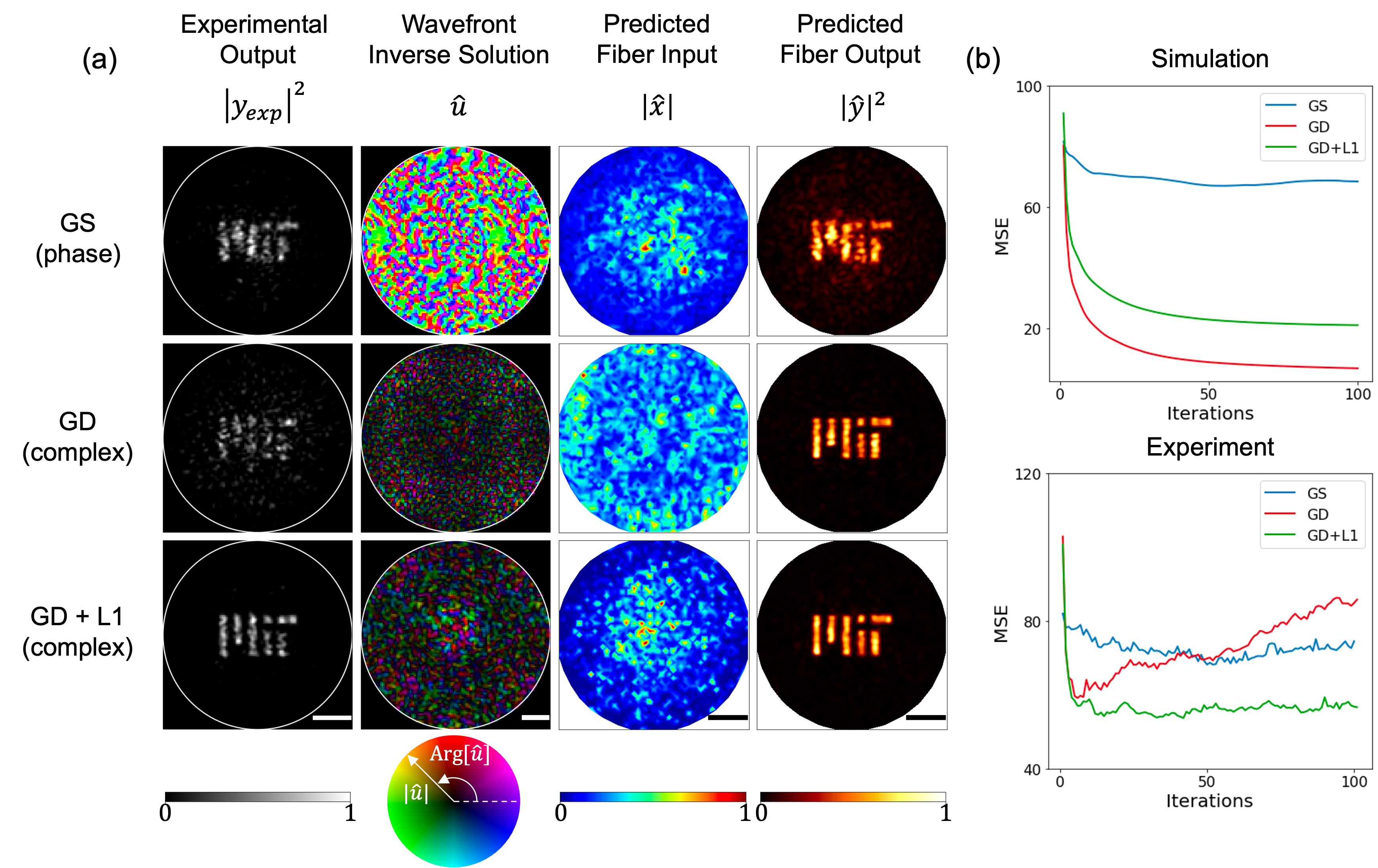}
\caption{\textbf{Analysis of the projection performance of the sparsity-constrained optimization.} (a) Simulation and Experimental results of projection through a graded-index multimode fiber (GIF50C, Thorlabs). (b) Optimization curves in simulation and experiment. The sparsity-constrained method (GD + L1) achieves consistent high-quality fiber output in simulation and experiment, which is attributed to the sparse representation it finds in the random basis provided the complex medium (i.e., the predicted fiber input in (a)). The scale bars for experimental outputs, predicted fiber inputs, and predicted fiber outputs are \SI{10}{\micro\meter}, and the scale bar for wavefront inverse solutions is \SI{1}{\milli\meter}.}\label{figure3}
\end{figure*}

By leveraging the sparse-to-random transformation in complex media, we propose a sparsity-constrained optimization framework which incorporates the dimensionality limitation of devices in the form of $l_1$ regularization (Fig. \ref{figure1}(b) and Methods \ref{subsec-optimization}). We test the performance of the sparsity-constrained optimization framework in the experimental setup shown in Supplementary Figure S5 (see Methods \ref{subsec-TM} for more details) using a graded-index multimode fiber (GIF50C, Thorlabs) as the complex medium, and compare it with two commonly used methods: Gerchberg-Saxton (GS) algorithm \cite{Gerchberg1972, Cizmar:11} with phase-only constraint at the conjugate plane of the DMD, and gradient descent (GD) method \cite{Zhang:17, Lee2022_dmd} without constraints. The implementation details of three methods are provided in Supplementary Note 3 and 4. Sparsity-constrained optimization consistently outperforms the GS algorithm and the GD method for various targets (Fig. \ref{figure3}(a) and Fig. \ref{figure4}), achieving considerably higher-fidelity light manipulation through the fiber (higher accuracy and better speckle suppression, more see the next section and Table \ref{tab_psnr}, \ref{tab_wc}, \ref{tab-fashionmnist}).

Here we dive in the graded-index multimode fiber experiment to gain quantitative insights into why sparsity-constrained optimization yields considerably better results and why this can be potentially extended to almost any existing DMD-based wavefront shaping systems, different complex media types, and different light manipulation target pools. The root cause of the improvement can be dissected from three perspectives. First, from the perspective of compressive sampling, generic patterns are compressible in a random basis \cite{Candes2006, Candes2008}, and complex media provides a natural platform as an analogue randomizing compressor \cite{Liutkus2014} (more examples see Supplementary Figure S2). Thus, transforming a wavefront by a random matrix together with $l_1$ minimization is an effective compression strategy for a wide variety of patterns through complex media.

Secondly, from the perspective of DMD shaping fidelity, the inherent limitation of DMDs can be approximated by the number of non-zero modes in the Fourier domain (Fig. \ref{figure2}(b-c) and Supplementary Note 1), which fits well with the $l_1$ minimization of the compression strategy. Our objective function leverages this synergy between the sparse representation of solutions and the dimensionality limitation of wavefront shaping. The random property of complex media allows the conversion of a high-bandwidth signal to a low-bandwidth one in a random basis with a sampling rate far below the Nyquist sampling rate. Such sparsity perfectly remedies the dimensionality limitation of DMDs, and these two properties can be seamlessly combined and efficiently solved in the shared domain of the Fourier plane of the DMD and the input plane of scattering media. Such synergy results in better solution optimality (i.e., wavefront shaping fidelity) and better solution feasibility (i.e., the consistency between the predicted outputs and the experimental outputs), as presented in Fig. \ref{figure3}(a). A comprehensive discussion on the tradeoff between the solution feasibility and optimality is entailed in Supplementary Note 5.

Thirdly, from the perspective of robustness and practicality, compared to other constraints (e.g., phase-only, or binary), $l_1$ minimization has the theoretical guarantee that it can stably and accurately reconstruct nearly sparse signals from dramatically undersampled data in an incoherent domain \cite{Candes2008}. Fig. \ref{figure3}(b) illustrates the robustness of this sparsity-constrained optimization method, of which the mean squared error (MSE) curves consistently descend and converge in both the simulation and the experiment (more see Supplementary Note 5). Besides the robustness, this simple objective function also has a closed-form expression of its gradient, which makes the computation time advantageous compared to existing methods and compatible with the hardware implementation (Supplementary Figure S10). More detailed analyses and descriptions can be found in Supplementary Note 6.

\subsection{Extension to different complex media}\label{subsec4}
\begin{figure*}[h]
\centering
\includegraphics[width=1.0\textwidth]{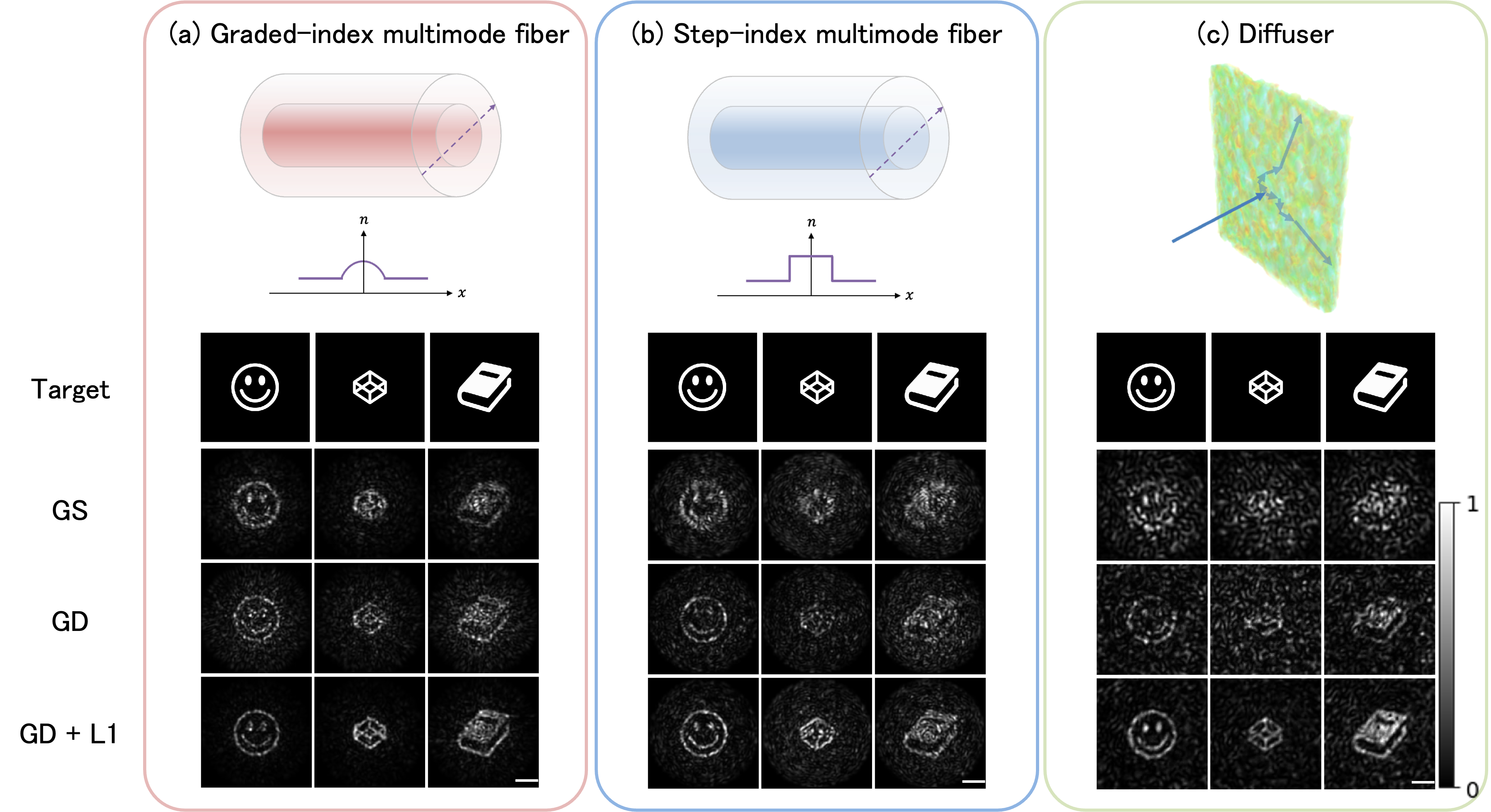}
\caption{\textbf{Experimental demonstration of enhanced projection quality in different complex media.} (a) Graded-index multimode fiber. (b) Step-index multimode fiber. (c) Diffuser. The scale bars are \SI{10}{\micro\meter}.}\label{figure4}
\end{figure*}

To show its broad applicability, we test our method on different types of complex media, including graded-index multimode fibers (GIF50C, Thorlabs), step-index multimode fibers (FG050LGA, Thorlabs), and diffusers (DG10-600, Thorlabs). We use the same experimental setup for all three experiments, as shown in Supplementary Figure S5. The experimental TM of each medium is characterized as described in Methods \ref{subsec-TM}. After calibrating the TMs, we optimize the projections through each medium using the three methods described in Section \ref{subsec3}. The implementation details of the optimization are provided in Supplementary Note 3. Despite differences in the underlying scattering mechanisms, the proposed method demonstrates consistent improvements in the projection quality in Fig. \ref{figure4} and Table \ref{tab_psnr} by leveraging the sparse-to-random transformation exhibited in all three media to overcome the dimensionality limitation of wavefront shaping. The proposed method shows a significant improvement of up to 2.73 dB in peak-to-noise-ratio (PSNR), which translates to an 89\% enhancement in projection accuracy. A complete panel of the projected images in Table \ref{tab_psnr} can be found in Supplementary Figure S11. To characterize the effectiveness of the proposed method in reducing background speckle noise in the projected images, we calculate the speckle suppression in Table \ref{tab_wc}, which is defined as the ratio of the light intensity of the foreground and background. It is also referred to as Weber contrast in visual perception and imaging processing \cite{Peli:90}. The proposed method achieves a remarkable enhancement of speckle suppression up to 126\% compared to the other two methods. Such consistent improvement in different types of complex media demonstrates the generalizability of the proposed method, and it can be readily adopted for various light manipulation projects involving complex media.

\begin{table}[h]
\footnotesize
\begin{center}
\begin{minipage}{\textwidth}
\caption{Comparison of projection fidelity. Peak signal-to-noise ratio (PSNR, first row values for each target) and multi-scale structural similarity (MS-SSIM, second row values for each target) are shown for different targets. GI-MMF: graded-index multimode fiber. SI-MMF: step-index multimode fiber.}\label{tab_psnr}
\begin{tabular*}{\textwidth}{@{\extracolsep{\fill}}cccccccccc@{\extracolsep{\fill}}}
\toprule%
& \multicolumn{3}{@{}c@{}}{GI-MMF} & \multicolumn{3}{@{}c@{}}{SI-MMF} & \multicolumn{3}{@{}c@{}}{Diffuser} \\\cmidrule{2-4}\cmidrule{5-7}\cmidrule{8-10}%
Target & GS & GD & GD + L1 & GS & GD & GD + L1 & GS & GD & GD + L1 \\
\midrule
\multirow{4}{*}{\raisebox{-.5\height}{\includegraphics[width=0.07\textwidth, height=0.07\textwidth]{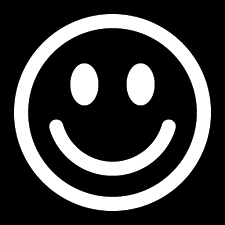}}} \\ & 16.17 & 14.99 & \textbf{16.80} & 16.22 & 15.94 & \textbf{17.86} & 15.51 & 15.41 & \textbf{17.90}\\
& 0.65 & 0.50 & \textbf{0.72} & 0.62 & 0.53 & \textbf{0.68} & 0.47 & 0.39 & \textbf{0.60}\\[.5mm]\\
\multirow{4}{*}{\raisebox{-.5\height}{\includegraphics[width=0.07\textwidth, height=0.07\textwidth]{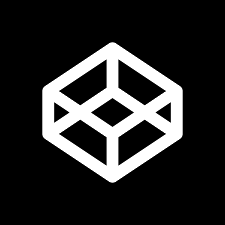}}} \\ & 17.63 & 16.78 & \textbf{19.42} & 17.30 & 17.02 & \textbf{18.81} & 16.36 & 15.86 & \textbf{19.13}\\
& 0.70 & 0.57 & \textbf{0.80} & 0.57 & 0.49 & \textbf{0.64} & 0.39 & 0.30 & \textbf{0.58}\\[.5mm]\\
\multirow{4}{*}{\raisebox{-.5\height}{\includegraphics[width=0.07\textwidth, height=0.07\textwidth]{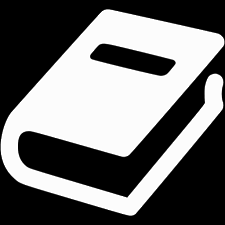}}} \\ & 14.54 & 13.53 & \textbf{15.47} & 13.81 & 13.50 & \textbf{14.90} & 13.87 & 13.36 & \textbf{16.94}\\
& 0.58 & 0.48 & \textbf{0.67} & 0.48 & 0.44 & \textbf{0.54} & 0.41 & 0.37 & \textbf{0.59}\\[.5mm]\\
\multirow{4}{*}{\raisebox{-.5\height}{\includegraphics[width=0.07\textwidth, height=0.07\textwidth]{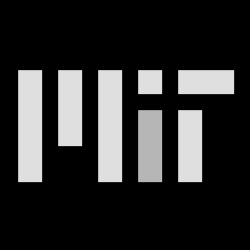}}} \\ & 16.80 & 15.43 & \textbf{17.99} & 16.60 & 15.61 & \textbf{18.10} & 17.04 & 16.89 & \textbf{20.52}\\
& 0.66 & 0.48 & \textbf{0.73} & 0.59 & 0.50 & \textbf{0.68} & 0.53 & 0.48 & \textbf{0.75}\\[.5mm]\\
\multirow{4}{*}{\raisebox{-.5\height}{\includegraphics[width=0.07\textwidth, height=0.07\textwidth]{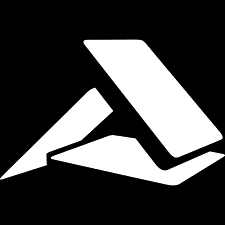}}} \\ & 17.27 & 15.25 & \textbf{18.17} & 16.75 & 15.84 & \textbf{17.70} & 15.78 & 15.09 & \textbf{18.58}\\
& 0.68 & 0.49 & \textbf{0.75} & 0.57 & 0.48 & \textbf{0.62} & 0.44 & 0.36 & \textbf{0.63}\\[.5mm]\\
\multirow{4}{*}{\raisebox{-.5\height}{\includegraphics[width=0.07\textwidth, height=0.07\textwidth]{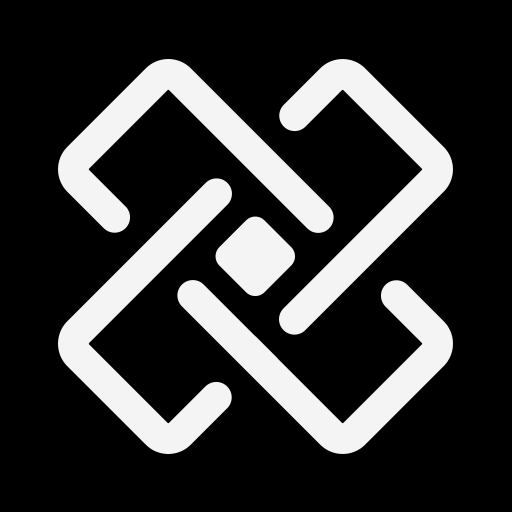}}} \\ & 16.20 & 14.54 & \textbf{17.00} & 16.52 & 16.00 & \textbf{18.23} & 15.25 & 14.60 & \textbf{17.27}\\
& 0.67 & 0.44 & \textbf{0.71} & 0.67 & 0.56 & \textbf{0.75} & 0.41 & 0.38 & \textbf{0.60}\\
[.5mm]\\
\midrule
\multirow{2}{*}{Avg.} & 16.44 & 15.09 & \textbf{17.48} & 16.20 & 15.65 & \textbf{17.60} & 15.63 & 15.20 & \textbf{18.39}\\
& 0.66 & 0.50 & \textbf{0.73} & 0.58 & 0.50 & \textbf{0.65} & 0.44 & 0.38 & \textbf{0.62}\\
\botrule
\end{tabular*}
\end{minipage}
\end{center}
\end{table}

\begin{table}[h]
\footnotesize
\begin{center}
\begin{minipage}{\textwidth}
\caption{Comparison of speckle suppression (quantified as Weber contrast). The average Weber contrast values are computed using the targets in Table \ref{tab_psnr}.}\label{tab_wc}
\begin{tabular*}{\textwidth}{@{\extracolsep{\fill}}cccccccccc@{\extracolsep{\fill}}}
\toprule%
& \multicolumn{3}{@{}c@{}}{GI-MMF} & \multicolumn{3}{@{}c@{}}{SI-MMF} & \multicolumn{3}{@{}c@{}}{Diffuser} \\\cmidrule{2-4}\cmidrule{5-7}\cmidrule{8-10}%
 & GS & GD & GD + L1 & GS & GD & GD + L1 & GS & GD & GD + L1 \\
\midrule
WC\footnotemark & 9.47 & 6.13 & \textbf{15.01} & 4.68 & 4.18 & \textbf{7.19} & 4.09 & 4.66 & \textbf{9.24}\\
\botrule
\end{tabular*}
\footnotetext[1]{Weber contrast (WC) is defined as $\frac{I-I_b}{I_b}$ \cite{Peli:90}, where $I$ and $I_b$ are the luminance of the foreground and the background, respectively.}
\end{minipage}
\end{center}
\end{table}

\subsection{Extension to different target patterns}\label{subsec-application}
\begin{figure*}[h]
\centering
\includegraphics[width=0.7\textwidth]{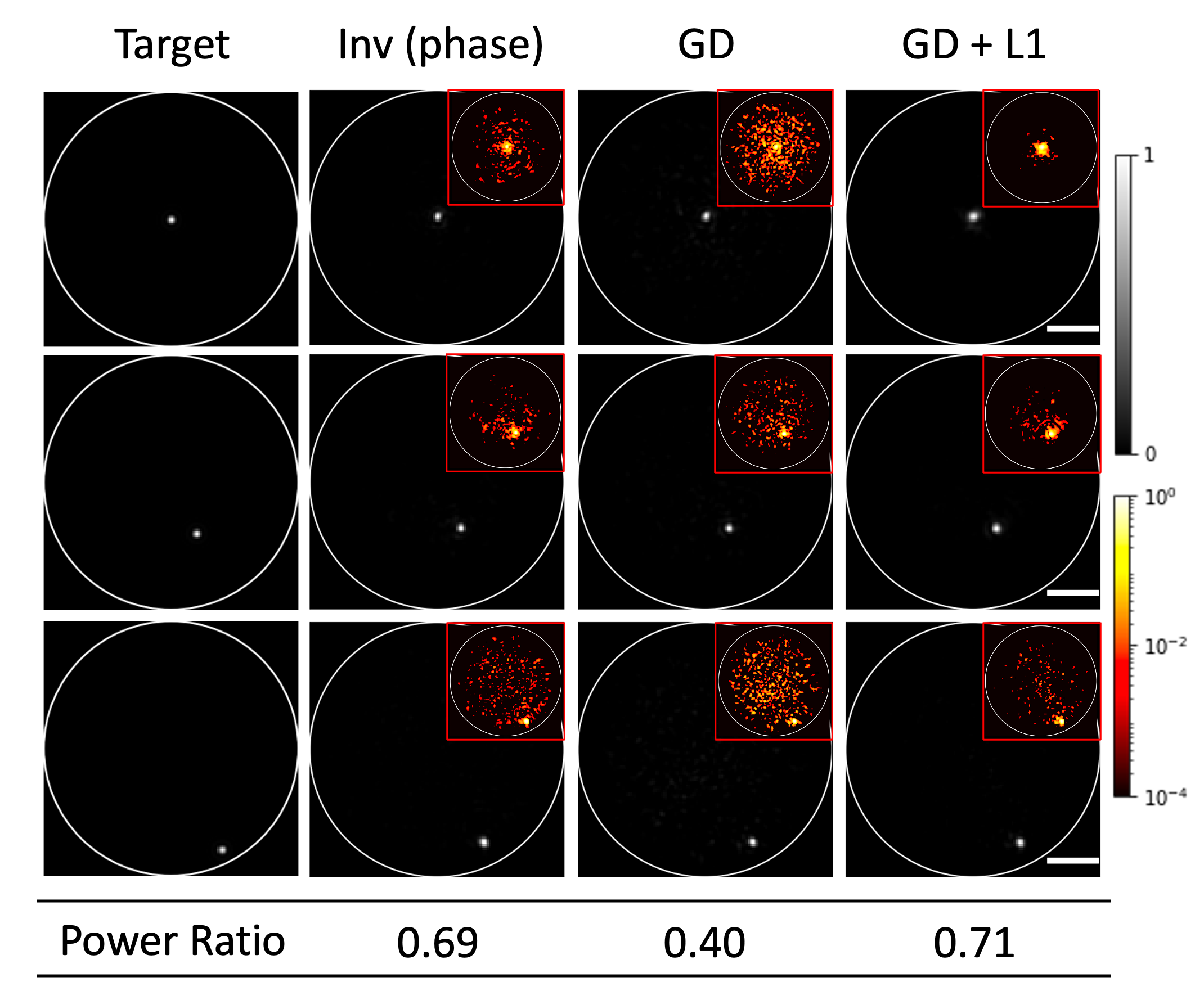}
\caption{\textbf{Diffraction-limited foci through a graded-index multimode fiber.} Three exemplary foci at different locations are demonstrated using different methods: Inv (phase): matrix inversion with phase-only wavefronts. GD: gradient descent method. GD + L1: sparsity-constrained optimization method. The insets are the corresponding log-scale images. The values are the average power ratio \cite{Gomes:22} of 77 foci on a grid with a \SI{5}{\micro\meter} spacing at the distal end. The scale bars are \SI{10}{\micro\meter}.}\label{figure6}
\end{figure*}

\begin{figure*}[h]
\centering
\includegraphics[width=1.0\textwidth]{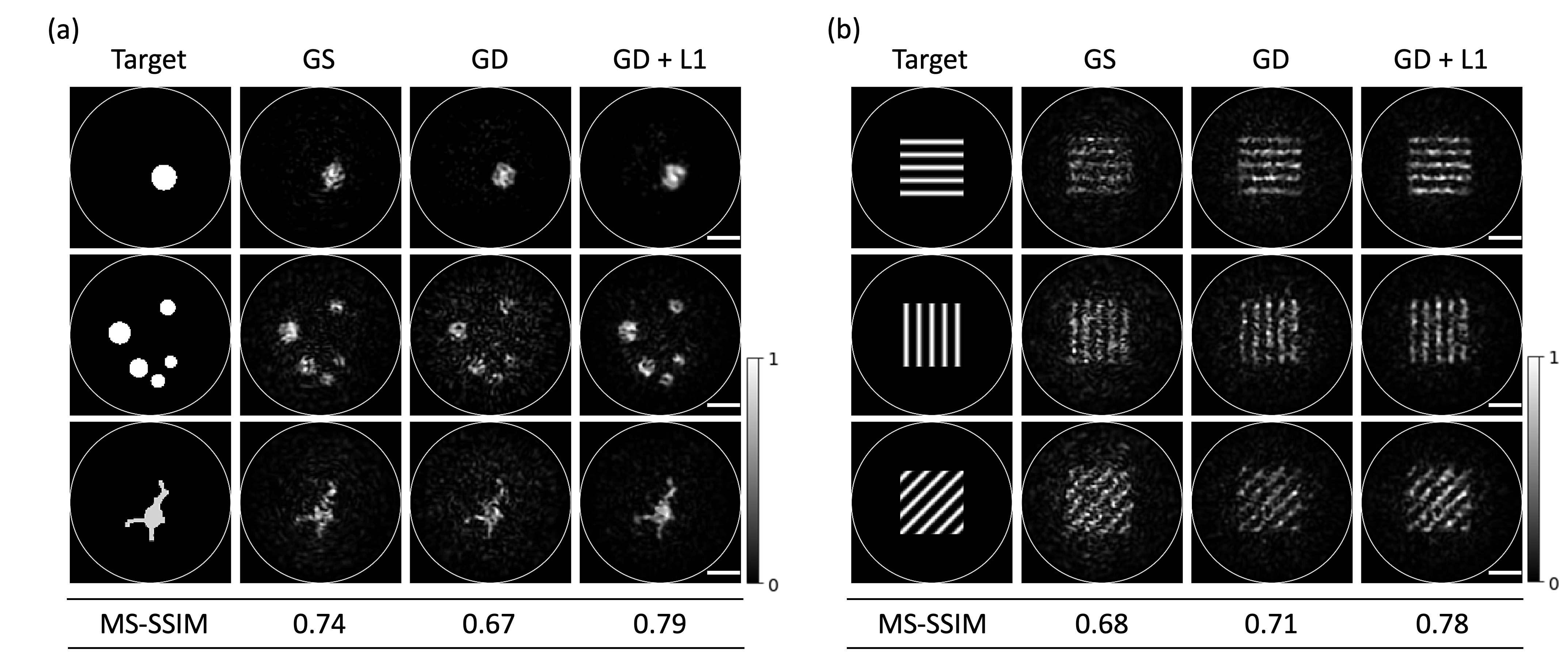}
\caption{\textbf{Potential applications of high-fidelity high-speed wavefront shaping through scattering} (a) Generation of masks for holographic optogenetics. (b) Generation of periodic patterns for compressive imaging (e.g., single-pixel imaging). The scale bars are \SI{10}{\micro\meter}.}\label{figure7}
\end{figure*}

\begin{figure*}[h]
\centering
\includegraphics[width=0.95\textwidth]{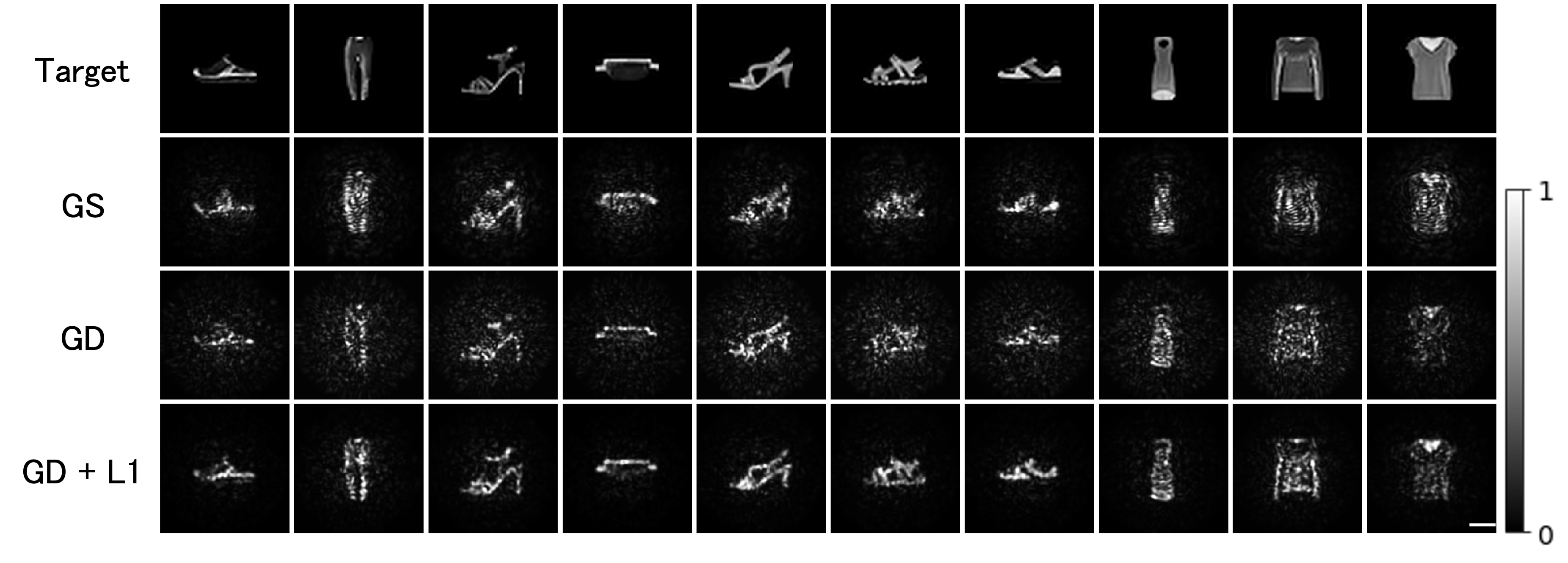}
\caption{\textbf{Experimental demonstration of gray-level image projection through scattering} The experiment involves testing 50 images extracted from the Fashion-MNIST dataset, with five examples from each of the ten categories. Ten exemplary projected images are presented. The scale bars are \SI{10}{\micro\meter}. Table \ref{tab-fashionmnist} displays the quantitative evaluation of the projected image quality.}\label{figure5}
\end{figure*}

In Fig. \ref{figure6}, we generate diffraction-limited foci through a graded-index fiber (GIF50C, Thorlabs), which is an important application in endoscopic imaging. Compared to standard matrix inversion method with phase-only conjugate wavefronts \cite{Turtaev:17}, our method effectively suppresses the residual field in the background region and achieves a slightly higher focusing efficiency. This is consistent with the discussion in Section \ref{subsec3} that our method effectively identifies the sparse and robust representations of generic patterns in the random basis. The relatively small improvement is owing to the already sparse solution in the random speckle basis, which is also the fiber input domain, for a diffraction-limited point as shown in Supplementary Figure S3. For a more complicated pattern of which the inverse solution is denser in the random speckle basis (e.g., the second row in Supplementary Figure S3), the effect of imposing a proper sparsity constraint in the optimization is substantial. In addition to diffraction-limited point focusing, we also demonstrate a wide variety of masks and patterns that can be applied in holographic optogenetics, compressive imaging, and optical communications through scattering, as illustrated in Fig. \ref{figure7} and Fig. \ref{figure5}. In all of the examples, our method achieves a consistent improvement throughout with a higher image contrast and lower speckle noise.

\begin{table*}[h]
\footnotesize
\begin{center}
\caption{Average PSNR and MS-SSIM of the Fashion-MNIST images that are projected using a graded-index multimode fiber. The values in the parentheses are the standard deviation.}\label{tab-fashionmnist}%
\begin{tabular}{@{}c|ccc@{}}
\toprule
Method & GS  & GD & GD + L1\\
\midrule
PSNR    & 18.36 (2.89) & 17.22 (2.95) & \textbf{19.30} (3.13) \\
MS-SSIM & 0.64 (0.09) & 0.55 (0.09) & \textbf{0.72} (0.10) \\
\botrule
\end{tabular}
\end{center}
\end{table*}

\section{Discussion}\label{sec3}
The presented results demonstrate that the proposed sparsity-constrained wavefront optimization framework can substantially enhance the projection quality across various types of complex media. It is worth noting that our method shares a resemblance with compressive sensing using random matrices together with $l_1$ minimization for sparse signal recovery \cite{Candes2008}. Complex media with intrinsic random property serve as a natural randomizer \cite{Liutkus2014} that allows for dimensionality reduction of patterns in a random basis. Finding this sparse representation is the key to achieving high-fidelity wavefront shaping with limited degrees of freedom. Our method successfully leverages the intrinsic dimensionality limitation of DMDs and pattern compression by the sparse-to-dense transformation through complex media, yielding substantially enhanced projection fidelity with single-shot wavefront shaping. Subsequently, we discuss the advantages of our method, followed by its potential limitations, in the context of real-world applications.

Firstly, our method only requires one frame per pattern, eliminating the need to compromise frame rate for projection fidelity. This allows for a full DMD frame rate of \SI{22}{kHz}, representing a ten-fold speed-up compared to temporal multiplexing methods and at least a 50-fold speed-up compared to LC-SLMs. Alternatively, for applications that do not require such a high frame rate, the excess frame rate can be leveraged for extended projection depth in microscopy \cite{Cao2023} or higher spectral resolution in hyperspectral imaging \cite{Dong:19}.
Besides the high-speed hardware operation, our method has the advantage of low computational complexity. Due to the simple form of the objective function, the gradient calculation is straightforward and the iterations can be much less computationally demanding than alternative iterative methods (e.g., phase retrieval algorithms, binary optimization, and neural networks). By directly coding the closed-form expression of the gradient as derived in Supplementary Note 4, the entire computation becomes a series of matrix multiplication, which is highly parallelizable in GPUs. As the benchmark of the computing time provided in Supplementary Note 6 and Figure S10, it only takes \SI{0.5}{\second} to optimize 1000 target patterns with a low-end GPU, largely attributed to the simplicity of the objective function (Equation \ref{eqn:lossfunction}). With such real-time optimization speed, our method is a promising solution to applications that involve dynamic complex media and require real-time optimization.

In addition, our method has wide applicability because it builds on two intrinsic properties - the dimensionality limitation of DMDs and the sparse-to-random transformation through complex media. These features make our method a versatile tool that can be used in combination with other methods. For instance, the sparsity constraint can be incorporated into temporal multiplexing methods to reduce the number of required frames for complicated 2D patterns or even 3D volumetric structures. End-to-end and deep-learning-based methods can benefit from the sparsity constraint, which may simplify the model complexity and decrease data dependence by providing a reliable physics prior. Furthermore, the same strategy can be applied to different devices or system setups that has a sparsity constraint due to limited degrees of freedom, and different types of complex media that randomize the signals. As a consequence, the proposed sparsity-constrained optimization framework and other approaches can work synergistically to boost performance in various applications that demand light manipulation in real-time with high accuracy.

While our method has shown promising results, there is still room for further enhancement. Firstly, the sparsity constraint in our approach approximates the wavefront error caused by limited amplitude modulation depth, but a real wavefront error also depends on the distribution of its angular spectrum as discussed in Supplementary Note 1. To improve the accuracy, one possible approach is to design a wavefront loss function that considers the effect of spectrum distribution. Despite this limitation, the sparsity constraint has a strong merit in its simplicity, offering an elegant understanding of the dimensionality limitation of spatial light modulation devices and ease of implementation. Secondly, the method is based on the characterization of the transmission matrix, which is susceptible to perturbations especially in a dynamic system. With the recent advancement in the compressive sampling techniques based on memory effect \cite{Li2021, Li2021_comp, Yilmaz2019, Chen:22} and single-ended calibration techniques based on reciprocity-induced symmetry \cite{Lee2020}, the characterization process can be considerably accelerated. Lastly, compared to phase-only constraints, one inherent drawback of allowing complex wavefront solutions in optimization is the lower power efficiency due to the rejection of partial incident light on the DMD. This can be potentially mitigated by a higher laser power since DMDs have a high damage threshold.

In conclusion, our investigation into the dimensionality limitation of wavefront shaping and the availability of sparse wavefront representations through complex media has successfully yielded a new method that enables unprecedented high-speed and high-fidelity projection through complex media. The proposed method and its results will benefit emerging, yet technologically challenging applications, such as non-invasive deep brain calcium imaging, high-speed holographic optogenetics, and miniaturized fiber-based 3D printing devices. We expect that our investigation into dimensionality limitation of wavefront shaping and the sparse and robust wavefront representations enabled by complex media will facilitate light manipulation within complex media in a wide range of existing systems involving DMDs and SLMs and serve as a catalyst for more systematic and universal approaches to tackling light scattering problems.

\section{Methods}\label{sec4}
\subsection{Sparsity-constrained wavefront optimization}\label{subsec-optimization}
As shown schematically in Fig. \ref{figure1}(b), our proposed sparsity-constrained wavefront optimization method seeks to solve for an inverse solution $\hat{x}$ that minimizes the loss function $\mathcal{L}$ given a desired projection pattern $y_t$:

\begin{equation}
    \hat{x} = \underset{x}{\arg\min} \ \mathcal{L}(x;y_t),
\label{eqn:sparsityOptimization}
\end{equation}
The loss function $\mathcal{L}$ consists of two terms: a data fidelity term that penalizes the difference between the target projection pattern $y_t$ and the pattern $y$ estimated by the forward model $y=Tx$, and a physics prior term that represents the sparsity constraint through $l_1$ regularization, also known as LASSO regression \cite{Tibshirani1996}:

\begin{equation}
    \mathcal{L}(x;y_t) = \frac{1}{2} \| y_t - \lvert Tx \rvert \|^2_2 + \lambda \| x \|_1,
\label{eqn:lossfunction}
\end{equation}
The introduction of $l_1$ regularization tends to suppress the coefficients of the less representative features to zero. The resulting solution falls onto a low-dimension manifold and therefore has a sparse representation. This optimization problem can be solved using the gradient descent method, and the inverse solution $\hat{x}$ can be used to obtain the estimated wavefront $\hat{v}$ in the image plane by performing an inverse Fourier transform:

\begin{equation}
    \hat{v}(\xi, \eta) = \mathcal{F}^{-1}\{ \hat{x}(\xi', \eta') \},
\label{eqn:IFT}
\end{equation}
Here $(\xi, \eta)$ and $(\xi', \eta')$ are Cartesian coordinates in the image plane and in the Fourier plane, respectively. Lastly, the estimated wavefront is encoded as a binary DMD hologram using the Lee hologram method \cite{Mirhosseini:13}:

\begin{equation}
    \hat{u}(\xi,\eta) = \frac{1}{2} + \frac{1}{2}\sgn\left[\cos{(k_0 (\xi + \eta) - \phi(\xi, \eta)) - \cos{(w(\xi, \eta))} }\right],
\label{eqn:leehologram}
\end{equation}
where $\phi(\xi, \eta) = \Arg\left[ \hat{v}(\xi, \eta) \right]$ is the phase of the estimated wavefront, $w(\xi, \eta) = \arcsin{\left[ \frac{\lvert\hat{v}\rvert}{\lvert\hat{v}\rvert_{\max}} \right]}$ is the arcsine of the normalized amplitude, and $k_0$ is the modulated carrier frequency determining the angle of the first diffraction order.

\subsection{Simulation and Experimental Implementation of Lee hologram method}\label{subsec-simulation}
The simulations of binary Lee holograms demonstrated in Fig. \ref{figure2}(b), Fig. \ref{figure2}(d), and Supplementary Figure S2-3 are constructed using scalar diffraction theory. To compute the wavefront in the Fourier plane generated by a binary Lee hologram in the image plane, we perform two steps: 1) Fourier transform of the binary hologram, and 2) screening the field outside the aperture of the spatial filter centered at the first diffraction order in the Fourier domain. Our simulations use a binary hologram with 512 $\times$ 512 pixels and a superpixel size of 4 $\times$ 4 in the Lee hologram method.

To model the transmission of light through a complex medium characterized by a TM as shown in Fig. \ref{figure2}(d), we convert the simulated wavefront in the Fourier plane ($\mathcal{F}\{u(\xi, \eta)\}$) to the pixel-based input mode domain of the TM by calculating the overlap integral with each input pixel mode $\psi_i$:

\begin{equation}
    x_i = \iint \psi_i^*\mathcal{F}\{u\} dA,
\label{eqn:modeoverlap}
\end{equation}
where $x = [x_1, x_2, \ldots, x_n]^T$ is a vectorized input of the TM. In the pattern reconstruction simulation shown in Fig. \ref{figure2}(d), the solutions are obtained by selecting the input modes with the $M$ greatest absolute values of the coefficients and setting the remaining coefficients to zero.

The experimental setup of the Lee hologram method is depicted in Supplementary Figure S5. The standard configuration consists of a DMD and a 4$f$ system with a spatial filter located in the Fourier plane. We employ the same parameters for the number of pixels and the size of superpixels as in our simulation. In our setup, we use an objective (OBJ1) to couple the wavefront into the complex medium. To observe the wavefront generated by the Lee hologram method in the Fourier plane, we remove the complex medium shown in Supplementary Figure S5 and adjust two objectives to be confocal. To examine the image projection through complex media, we use the same setup as depicted in Supplementary Figure S5.

\subsection{Experimental setup and characterization of transmission matrix}\label{subsec-TM}
In the experimental setup depicted in Supplementary Figure S5, a 100-mW, 488-nm continuous-wave laser (Sapphire 488 SF NX, Coherent) is utilized for illumination. The laser beam is expanded by a 4$f$ system (L1 and L2) with 10$\times$ magnification to match a circular region of \SI{7}{\centi\meter} in diameter, equivalently 512 pixels, on the DMD (V-7001, Vialux). The Lee hologram method is applied to generate a predefined complex wavefront in the first diffraction order, and the other diffraction orders are blocked with a spatial filter (SF) in the Fourier plane. An objective (RMS20X, Olympus) is used to focus the wavefront onto the input plane of a complex medium, and another objective (RMS10X, Olympus) collects the resulting speckle on the output plane. The speckle image is formed on a monochrome camera (Mako G-040B, Allied Vision) after passing through a 4$f$ system (OBJ2 and L5) with 16.7$\times$ magnification. For the fibers used in the experiments (GIF50C and FG050LGA, Thorlabs), the length is approximately \SI{15}{\centi\meter}.

To determine the transmission matrix of the complex medium, we perform raster scanning at the proximal end in the Fourier plane of the DMD and acquire the corresponding complex-field speckles at the distal end \cite{Popoff2010, Turtaev:17}. For each of the complex media used in the experiments, we scan 1941 foci with a spacing of \SI{1.0}{\micro\meter} across a circular region with a diameter of \SI{50}{\micro\meter}. To achieve a diffraction-limited beam during raster scanning, we calibrate the wavefront aberration caused by the DMD using Zernike polynomials of $20^{th}$ order. The resulting speckles are split into two orthogonal linear polarization states by a beam displacer and measured using off-axis holography. To reduce phase instability, we measure a reference speckle to characterize and compensate the temporal phase variation caused by environmental vibration. Finally, we combine the two sub-matrices associated with the two polarization states in the output to generate the transmission matrix.

\bmhead{Code Availability} The codes for the sparsity-constrained wavefront optimization are available at \url{https://www.dropbox.com/sh/nvdcug1il0k7n4m/AADDQFonIwChJEDmQpywTYLWa?dl=0}.
\bmhead{Data Availability} The data that support the findings of this study are available from the corresponding author upon reasonable request.
\bmhead{Acknowledgments} We acknowledge the support from Jameel Clinic, Scialog, and MIT EECS and RLE startup funds. We would like to express our sincere gratitude to Martin Villiger for his valuable insights and guidance on the optical system setup and technical issues. We also extend our appreciation to Kristina Monakhova for providing constructive comments on an earlier version of the manuscript that greatly improved its clarity and quality.

\bmhead{Author contributions} L.Y. and S.Y. conceived the idea of the project. S.Y. supervised the research. L.Y. built the optical setup and performed the wavefront shaping experiments and simulations. L.Y. and S.Y. wrote the manuscript.

\bmhead{Competing interests} The authors declare no competing interests.

\bmhead{Supplementary information} A supplementary file is provided to accompany this manuscript, which contains additional data and analyses.

\bibliography{bibliography}



\section*{Supplementary material}

# High-Fidelity and High-Speed Wavefront Shaping by Leveraging Complex Media:
## Supplementary Information

Li-Yu Yu[1] and Sixian You[1]

[1] *Department of Electrical Engineering and Computer Science, Massachusetts Institute of Technology, Cambridge, 02142, Massachusetts, USA*

## Supplementary Note 1: Sparsity constraint in the Fourier plane

To understand the Fourier-domain characteristics of the wavefront produced by a spatial light modulation device with limited amplitude modulation depth, we begin by considering an arbitrary wavefront in the image plane decomposed into a superposition of multiple planewaves:

$$f(\mathbf{x}) = \sum_{n=1}^{N} \alpha_n e^{j\mathbf{k}_n \cdot \mathbf{x}}, \tag{S1}$$

where $N$ is the total number of planewaves, and $\alpha_n$ and $\mathbf{k}_n$ are the amplitude and the wavevector of each planewave. To measure the discrepancy between an ideal wavefront and the actual wavefront produced by a spatial light modulation device that lacks amplitude modulation capability (e.g., a phase-only SLM), we can compare the ideal intensity distribution with a uniform distribution:

$$|f(\mathbf{x})|^2 = \left( \sum_{n=1}^{N} \alpha_n e^{j\mathbf{k}_n \cdot \mathbf{x}} \right) \left( \sum_{m=1}^{N} \alpha_m^* e^{-j\mathbf{k}_m \cdot \mathbf{x}} \right)$$
$$= \sum_{n=1}^{N} |\alpha_n|^2 + \sum_{r \in \mathcal{R}} \alpha_r \cos\left( \Delta\mathbf{k}_r \cdot \mathbf{x} + \phi_r \right), \tag{S2}$$

where

$$\mathcal{R} = \{(n, m) \mid n < m\}, \tag{S3a}$$
$$\alpha_r = 2|\alpha_n^{(r)} \alpha_m^{(r)*}|, \tag{S3b}$$
$$\Delta\mathbf{k}_r = \mathbf{k}_n^{(r)} - \mathbf{k}_m^{(r)}, \tag{S3c}$$
$$\phi_r = \arg\left( \alpha_n^{(r)} \alpha_m^{(r)*} \right), \tag{S3d}$$

The first term in Equation S2 is the mean intensity and can be generated using a uniform phase-only wavefront. The second term represents the wavefront mismatch, which arises from the interference between different planewaves. This mismatch can be quantified by computing the intensity variance:

$$Var(|f(\mathbf{x})|^2) = \frac{1}{A} \int_\Omega \left[ \sum_{r \in \mathcal{R}} \alpha_r \cos\left(\Delta\mathbf{k}_r \cdot \mathbf{x} + \phi_r\right) \right]^2 d\mathbf{x}$$

$$= \frac{1}{A} \int_\Omega \left[ \sum_{r \in \mathcal{R}} \alpha_r^2 \cos^2\left(\Delta\mathbf{k}_r \cdot \mathbf{x} + \phi_r\right) \right.$$

$$\left. + \sum_{r_1, r_2 \in \mathcal{R}, r_1 \neq r_2} \alpha_{r_1} \alpha_{r_2} \cos\left(\Delta\mathbf{k}_{r_1} \cdot \mathbf{x} + \phi_{r_1}\right) \cos\left(\Delta\mathbf{k}_{r_2} \cdot \mathbf{x} + \phi_{r_2}\right) \right] d\mathbf{x}, \quad \text{(S4)}$$

The region on the wavefront device that is actively modulated is represented by $\Omega$ and has an area of $A$. Equation S4 manifests the relationship between sparsity in the Fourier plane and wavefront fidelity. When a wavefront is dense in the Fourier plane, i.e., $N$ is large, the intensity variance is likely to be large, as the size of the collection $\mathcal{R}$ is quadratically proportional to $N$. It is important to note that the second term in Equation S4 can only persist when two wavevector mismatches are nearly identical ($\Delta\mathbf{k}_{r_1} \approx \Delta\mathbf{k}_{r_2}$). This situation can occur when a wavefront is highly structured in the Fourier plane, such as lines or uniform grid points, which can explain the poor projection fidelity of dense and uniform patterns, as shown in Fig. 2(c) and Fig. 2(e).

The above discussion also applies to a spatial light modulation device with a discrete amplitude modulation depth, e.g., DMDs using the Lee hologram method [1, 2]. In this case, the wavefront mismatch becomes the sum of variances corresponding to each discrete amplitude level. Although the calculation becomes more complicated and depends on the threshold values for each amplitude level, the sparsity constraint remains valid since it determines the number of terms that contribute to the mismatch, as shown in Equation S4. The result in Figure S1 verifies the decreasing fidelity with increasing density of the projected patterns in the Fourier plane of a DMD. The rapidly increasing mean squared error (MSE) for the patterns with more connected structures is consistent with the second term in Equation S4 as discussed.

Supplementary Note 1 provides a simplified theoretical explanation for our experimental observations by deriving the wavefront error as a result of the limited degrees of freedom of DMDs. More advanced theoretical model including complex field modulation can provide more exact insight. Nevertheless, since the ultimate control degrees of freedom are the number of pixels in DMD, our Note serves the goal of a general and intuitive mathematical explanation.

## Supplementary Note 2: Characterization of speckle properties for pattern compression in complex media

Here we provide the characterization of speckle properties and a more thorough explanation on how patterns can be compressible in complex media, followed by the applicability of our method in the context of real-world applications. We first calculate the field-field speckle correlation matrices [3] of the diffusers with 600 and 1500 grit polishes in Figure S4(a-b). The 600-grit diffuser shows weaker memory effect (i.e., weaker non-diagonal elements in the correlation matrix), which is in line with our empirical observations in Fig. 2(e) that a stronger scattering effect can facilitate wavefront shaping. In addition, we calculate the coherence of the transmission matrices in Figure S4(c), an imperative factor for sparse signal recovery in compressive sensing [4]. A low-coherence sensing matrix (e.g., a Gaussian i.i.d. random matrix) indicates a higher chance of recovering a sparse signal. The curves in Figure S4(a) concur with the theory that the diffuser with a low-coherence speckle basis (600-grit) achieves superior reconstruction.

To show that such level of scattering strength is relevant in real-world applications, we quantitatively characterize the diffuser following Ref [3]. The full-width at half-maximum (FWHM) and the full-width at

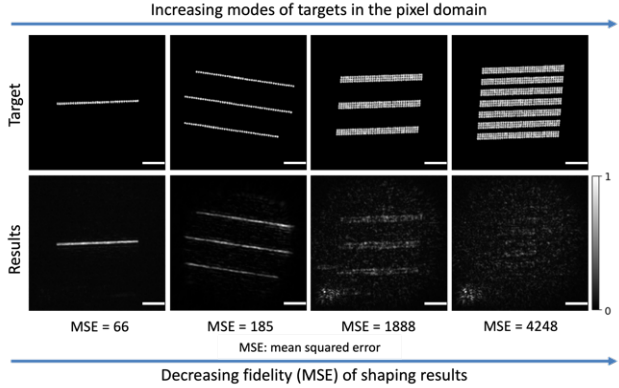

**Figure S1:** Experimental demonstration of the sparsity constraint in the Fourier plane. From left to right, the patterns with increasing density are generated in the Fourier plane of the DMD using the Lee hologram method. The scale bars are $10\,\mu m$.

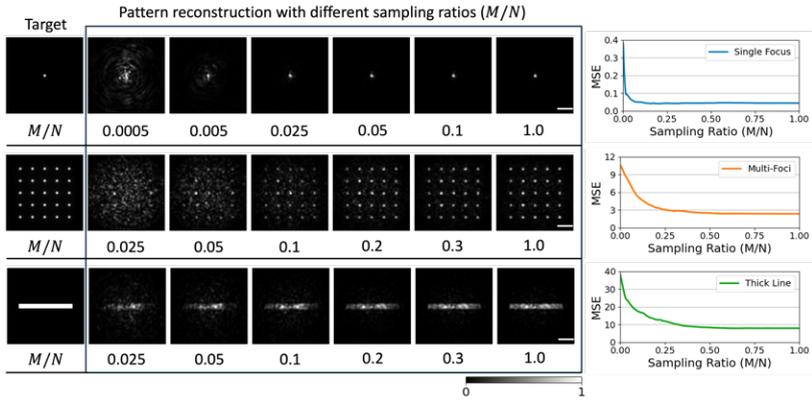

**Figure S2:** Simulations showing that patterns can be represented by a dramatically undersampled random basis (i.e., sparse representation in the random basis provided by the complex media). The patterns are reconstructed with different numbers of random modes in the simulation. The scale bars are $10\,\mu m$.

tenth-maximum (FWTM) of the correlation function (the inset in Figure S4(a)) of the diffuser are $1.5\,\mu m$ and $3.6\,\mu m$, respectively. Given the shorter wavelength used in our experiments ($488\,nm$), the scattering strength of the 600-grit diffuser is estimated to be on par with that of tissues with a $1\,mm$ thickness - a thickness commonly significant in bioimaging and optogenetics [5]. In this regime, the memory effect still persists, as evinced by the deviation of the curves between the diffusers and the ideal Gaussian i.i.d. random matrix. In other words, our method remains effective despite the presence of certain level of memory effect.

## Supplementary Note 3: Implementation of algorithms

The implementation details of the Gerchberg-Saxton (GS) algorithm and the gradient descent (GD) method are presented in this section.

The GS algorithm is a common phase-retrieval method to iteratively solve for a solution with unknown phase information [6, 7]. We adapt the GS algorithm, which follows the workflow illustrated in Figure S6. Starting from an initial random wavefront on the DMD plane, the forward model involves the a Fourier transform (FT) and a multiplication with the given transmission matrix ($T$). Then, a amplitude constraint defined by the target pattern is applied on the calculated output modes. The inverse model involves the reverse process of the forward model, including a multiplication with the Hermitian conjugate of the transmission matrix ($T^H$) and an inverse Fourier transform (IFT). Here, we approximate the inverse matrix by its Hermitian conjugate for better computational efficiency. The resulting wavefront of the inverse model will be constrained by the source amplitude, which is uniform in our case. The final binary Lee hologram is prepared after the algorithm stops at a predetermined number of iterations.

The GD method has been applied to a variety of wavefront shaping and computer-generated holography problems [8, 9] as an extremely tractable approach as long as the gradient of the problem is well-defined. We apply the GD method to solve for the gradient of Equation 2 in Section 4.1 with respect to the input modes ($x$), as the workflow shown in Figure S7. After the cacluation of the gradient, the input modes is updated with a predefined learning rate. After the iteration process is finished, the optimized input modes will be first transformed into the DMD plane and then converted into a binary Lee hologram. In practice, we implemented the GD method using *CuPy*, a lightweight *Python* library for GPU-accelerated computing. To speed up the optimization, we derived and coded the gradient explicitly. The derivation of the gradient is presented in Supplementary Note 4, and the performance on optimization time is entailed in Supplementary Note 6.

For target patterns without explicit phase information, the starting pattern was initialized in both algorithms by performing the GS algorithm with 10 iterations for a fair comparison. For target patterns with explicit phase information (e.g., a single focal point defined by Laguerre-Gaussian modes), the starting pattern was initialized by computing its phase-conjugated solution. We normalized the nuclear norm (i.e., the total power) of the transmission matrix to its number of columns (i.e., input modes) to avoid different scaling factors in different measurements. A standard ADAM optimization algorithm [10] was used to perform gradient descent, with a learning rate of 0.0004 or 0.004 for target patterns with or without phase information. The value of regularization parameter ($\lambda$) was chosen as 0.5 based on the empirical results, which has achieved high-fidelity across samples and complex media. The effect of different regularization parameters is illustrated in Figure S8. The results are not very sensitive to the choice of $\lambda$, showing a moderate change in PSNR between 0.4 to 2. The required number of iterations is determined by the convergence of the optimization curves. In Fig. 4, we performed 100 iterations for all methods to highlight the robustness of the proposed method. In the other experiments, 50 iterations were used.

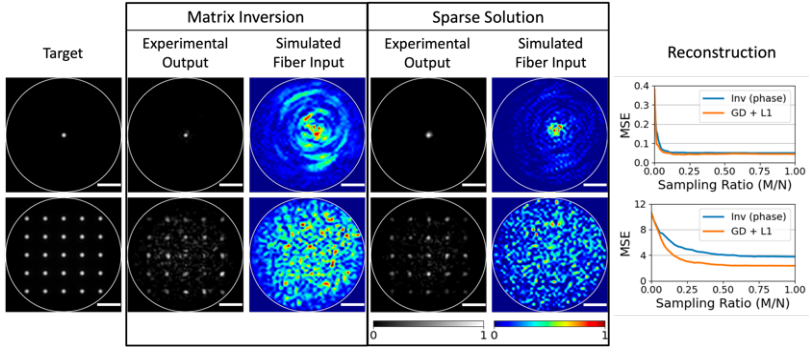

**Figure S3:** Comparison between the projection of patterns with different wavefront complexity in the random speckle basis. For the MSE curves, note the difference in saturation point and the difference in the improvement of $l_1$ minimization for the two patterns. The scale bars are $10\,\mu$m.

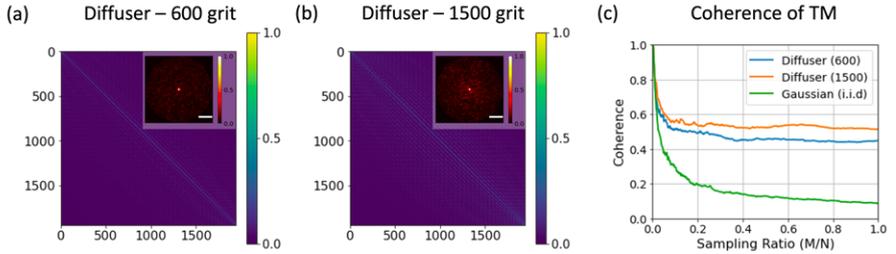

**Figure S4:** Analysis of speckle properties of the diffusers. (a-b) Field-field speckle correlation matrices of the diffusers with 600 and 1500 grit polishes, respectively. The insets are the speckle correlation maps with respect to the center input mode (c) Coherence of the experimental transmission matrices of the diffusers. The scale bars are $10\,\mu$m.

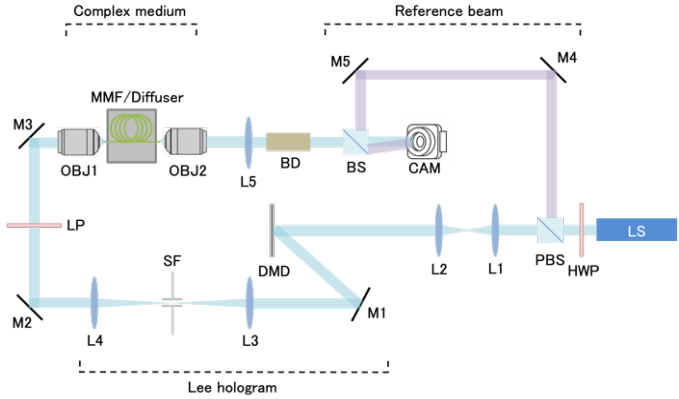

**Figure S5:** Experimental Setup. LS: sapphire laser at $488\,\text{nm}$. HWP: half waveplate. PBS: polarized beam splitter. L1-L5: lens with focal lengths of $30\,\text{mm}$, $300\,\text{mm}$, $150\,\text{mm}$, $150\,\text{mm}$, and $300\,\text{mm}$, respectively. M1-M5: mirrors. SF: spatial filter. LP: linear polarizer. OBJ1: objective with NA of 0.4. OBJ2: objective with NA of 0.25. BD: beam displacer. BS: beam splitter. CAM: camera.

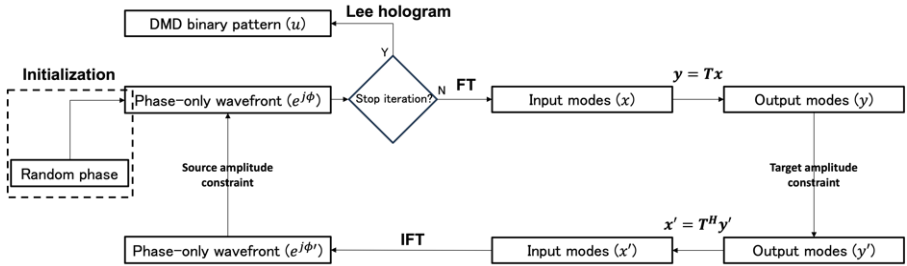

**Figure S6:** Workflow of the GS algorithm.

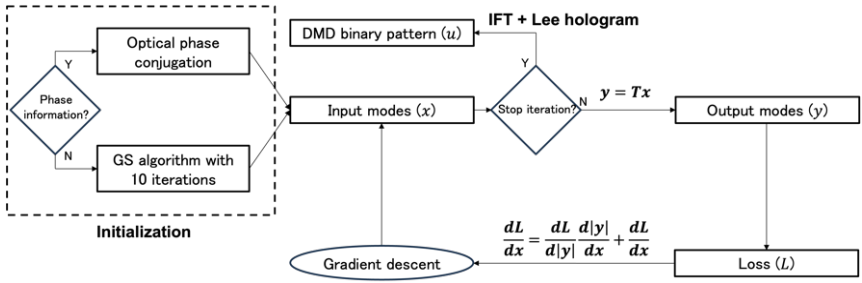

**Figure S7:** Workflow of the GD algorithm.

## Supplementary Note 4: Derivation of gradient

In this section, we derive the gradient of the loss function in Equation 2 explicitly for speeding up the optimization. To begin with, we decompose the gradient into two components:

$$\frac{dL}{dx} = \frac{\partial L}{\partial |y|} \frac{\partial |y|}{\partial x} + \frac{\partial L}{\partial x}, \tag{S5}$$

The first and the second components correspond to the derivative of the mean squared error and the regularization term, respectively. Each term in Equation S5 can be computed individually:

$$\frac{\partial L}{\partial |y|} = -|y_t|^T + |y|^T, \tag{S6a}$$

$$\frac{\partial |y|}{\partial x} = \frac{1}{2} \operatorname{diag}\left( y^* \odot \frac{1}{|y|} \right) T, \tag{S6b}$$

$$\frac{\partial L}{\partial x} = \frac{\lambda}{2} \left( x \odot \frac{1}{|x|} \right)^T, \tag{S6c}$$

where $\odot$ is Hadamard product and $\operatorname{diag}(v) = I_n \odot (v 1_n^T)$, which converts a vector to a diagonal matrix. Combining all the terms, we can obtain the closed-form expression of the gradient:

$$\frac{dL}{dx} = \frac{1}{2} \left[ \left( 1 - |y_t| \odot \frac{1}{|y|} \right) \odot y^* \right]^T T + \frac{\lambda}{2} \left( x \odot \frac{1}{|x|} \right)^T, \tag{S7}$$

The matrix formulation in Equation S7 indicates that it allows batch processing with computation complexity no harder than matrix multiplication.

## Supplementary Note 5: Solution optimality, feasibility, and numerical stability in wavefront shaping optimization

Here we expand the discussion on solution optimality, feasiblity, and numerical stability in Section 2.3, which are three factors that collectively influence the final projection quality through complex media.

Solution optimality describes the ability of an algorithm to converge to an optimal solution, assuming that the hardware achieves ideal wavefront shaping. It can be evaluated by the performance of the solution in the simulation (i.e., the predicted fiber outputs in Fig. 3(a)). On the other hand, solution feasibility describes the fidelity of the wavefront generated by the hardware given the numerical solution with unavoidable physical constraints. It can be evaluated based on the consistency between the results in the simulations and the experiments (i.e., predicted fiber outputs vs. experimental fiber outputs in Fig. 3(a)). We observe that the GS algorithm produces a highly feasible solution, which is reflected by the high consistency between the simulated and experimental fiber outputs. However, the resulting wavefront solution is overly simplistic due to the phase-only constraint, leading to poorer simulation and consequently poorer experimental results. These results and analyses indicate that the GS algorithm generates solutions with high feasibility but poor optimality due to over-constraint. On the other hand, while the GD method achieves a highly optimal solution in simulations, its experimental results are not as optimal due to the poor feasibility of the numerically optimized wavefront by the hardware. The highly complex wavefront solution and poor consistency between the simulated and experimental fiber outputs suggest that the GD method produces solutions with high optimality but poor feasibility due to a lack of constraints. This pattern becomes more apparent when we increase the number of iterations.

In addition to assessing feasibility and optimality, we also note a numerical instability problem caused by an ill-conditioned experimental TM. Inevitable noise in the TM measurement can become dominant when the overall light throughput is low, resulting in matrix inversion artifacts and unstable inverse solutions. The sparsity constraint in the form of $l_1$ regularization can help address this issue in the inverse problem, yielding inverse solutions with better numerical stability. By deliberately calculating the experimental mean squared error (MSE) curves and comparing them with the optimization curves in the simulation in Fig. 3(b), we observe an apparent overfitting issue in the GD method. By calculating the coupling efficiency (the fraction of input power coupled to the output computed as $diag(T^HT)$, where $H$ is the Hermitian conjugate of the matrix) in Figure S9(a), the noticeable differences in the distribution between each method shows up, especially in the lower row in Figure S9(a). This unbalanced coupling efficiency links to the distribution of the corresponding singular value spectra in Figure S9(b), indicating that the second transmission matrix is more ill-conditioned. The GS algorithm produces an input distribution that correlates with the input modes with higher coupling efficiency (the region inside the smaller dotted circle) as it computes $T^HT$ inherently during each iteration. In contrast, the GD method predicts an input distribution that correlates with the input modes with lower coupling efficiency (the region outside the larger dotted circle), as it solves the problem equivalently to performing matrix inversion $(T^HT)^{-1}$. The sparsity-constrained method estimates a more balanced input distribution across the regions of low and high coupling efficiency, showing strong resilience to the artifacts of the transmission matrix in an ill-conditioned inverse problem.

## Supplementary Note 6: Optimization time

Here we benchmarked the optimization time of our method and summarized the performance in Figure S10. As discussed in Supplementary Note 4, the closed-form expression of the gradient can be explicitly derived and hand-coded to accelerate the computation speed. Moreover, its matrix form allows parallel computing with GPUs for a substantial speed-up. With our customized implementation of gradient descent algorithm using *Python* package *CuPy* for GPU-accelerated computing on a personal computer with a 12-GB GPU (NVIDIA GeForce RTX 3080 Ti), it only took $0.5\,\mathrm{s}$ to perform the entire optimization for the wavefronts corresponding to 1000 desired patterns, as shown in Figure S10 (a). Such performance is similar to or even slightly faster compared to baseline methods. For example, the matrix inversion calculation for the same transmission matrix took $3.3\,\mathrm{s}$ to complete, while the optimization time for 1000 frames only cost $0.55\,\mathrm{s}$. Given that the matrix multiplication is highly parallelizable, additional speed-up can be expected by using a workstation-level GPU with larger memory and more computing cores. Based on these data and also the rapid development of parallel computing, the real-time computation for high-speed wavefront shaping with our method is achievable in the near future.

## Supplementary Note 7: Ablation study of system parameters

Here, we conducted an ablation study of the parameters of the wavefront shaping system to optimize our setup. In particular, we examined several combinations of the sizes ($d_{sp}$) and numbers ($N_{sp}$) of superpixels used in the Lee hologram method, as the number of variables could affect the performance significantly [11]. Following the procedures in Methods 4.3, we repeated the experiment in Section 2.3 using a graded index multimode fiber (GIF50C, Thorlabs) with different settings. In Figure S12, our method shows a consistent relative improvement compared to the other methods in all different settings. Such a consistent improvement is because our method is established upon two intrinsic properties: 1) dimensionality limitation of DMDs and 2) pattern compression via the sparse-to-random transformation through complex media. In fact, as our method optimizes upon the 1941 input modes ($x$ in Equation 2) defined by the transmission matrix instead of the pixels modes ($\hat{u}$ in Equation 4) on the DMD, it is robust to system parameters. In practice, we chose

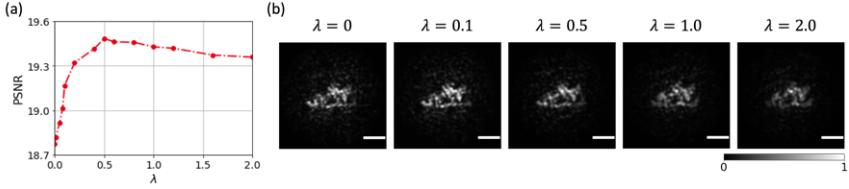

**Figure S8:** Ablation experiment of wavefront shaping with different regularization weights ($\lambda$). (a) The PSNR curve, and (b) a selected example of image projection with respect to different regularization weights. A graded-index multimode fiber (GIF50C, Thorlabs) was used in the experiment as the complex medium. The PSNR values were calculated by averaging the results of 50 examples randomly extracted from the Fashion-MNIST dataset. The scale bars are $10\,\mu m$.

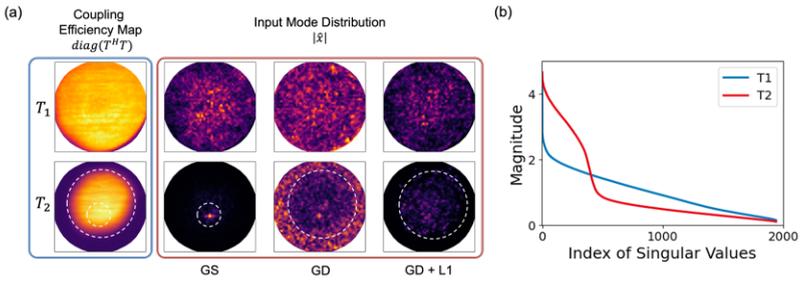

**Figure S9:** Numerical instability issue arising from the unbalanced coupling efficiency of the complex medium. (a) Comparison of experimental TMs with uniform and nonuniform coupling efficiency maps. The input mode distributions of the inverse solutions estimated by three different methods are shown: Gerchberg-Saxton (GS), gradient descent (GD), and gradient descent with sparsity constraint (GD + L1). The small and large dotted circles indicate the regions of high and low coupling efficiency, respectively. (b) Singular value spectra of the TMs in (a).

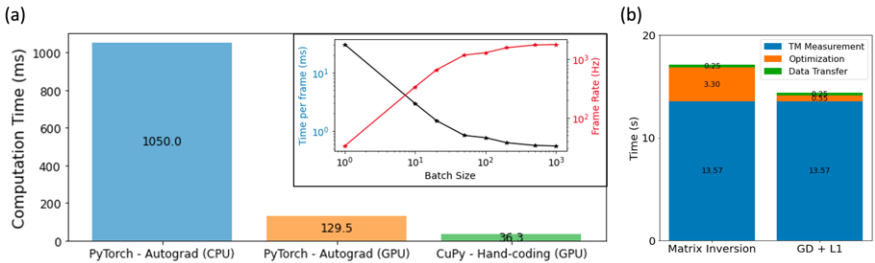

**Figure S10:** Benchmark of computation time of the sparsity-constrained optimization. (a) The computation time of a single frame with different implementations. The inset shows the optimization frame rate with different batch sizes using the customized hand-coding implementation. (b) Timing of the system operation for the characterization of a transmission matrix and wavefront shaping optimization on 1000 examples.

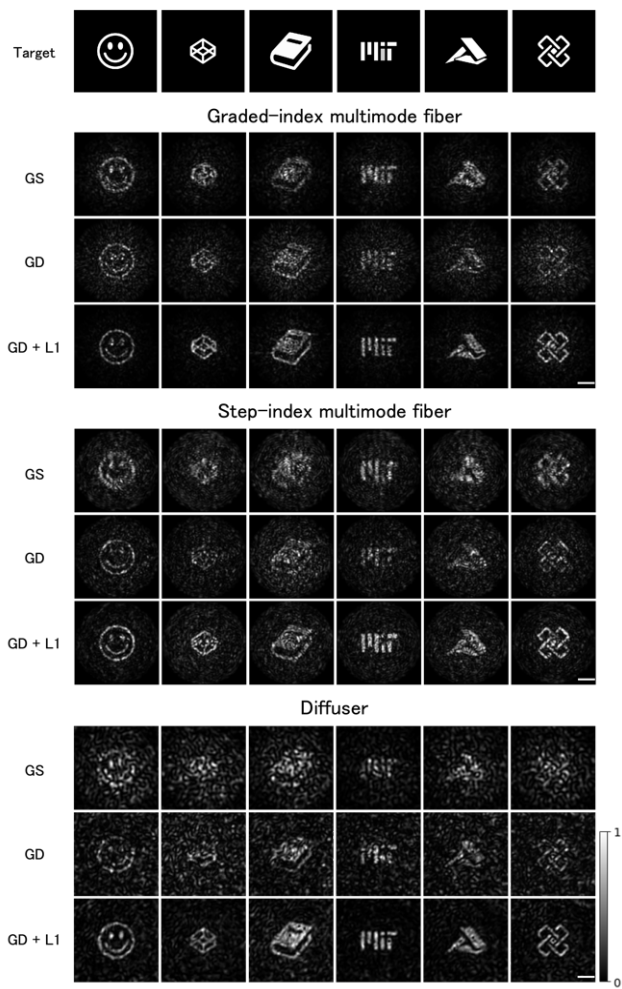

**Figure S11:** Experimental results of the projected images displayed in Table 1. The scale bars are $10\,\mu m$.

$d_{sp} = 4 \times 4$ and $N_{sp} = 128 \times 128$ in all the experiments elsewhere. We would like to note that the degraded performance for $d_{sp} = 8 \times 8$ is not owing to the insufficient number of superpixels but the insufficient spatial resolution, resulting in the occurrence of the aliasing effect in the Fourier plane.



| Parameters | | GS | GD | GD+L1 |
|---|---|---|---|---|
| $d_{sp}$ | $N_{sp}$ | MS−SSIM | | |
| 4×4 | 128×128 | 0.67 | 0.71 | **0.78** |
| 8×8 | 64×64 | 0.57 | 0.62 | **0.69** |
| 4×4 | 64×64 | 0.67 | 0.70 | **0.76** |

**Figure S12:** Wavefront optimization with different system parameters. The size and the number of super-pixels are denoted by $d_{sp}$ and $N_{sp}$, respectively. The scale bars are $10\,\mu$m.



\end{document}